\documentclass[twocolumn,prl,superscriptaddress,floatfix,aps]{revtex4}
\usepackage{bm}
\usepackage{epsfig}
\usepackage{graphicx}
\usepackage{amsfonts,amssymb,amsmath}
\begin{document}
\title{ Theory of Small Para-Hydrogen Clusters: Magic Numbers
and Superfluid Sizes}

\author{S.~A.~Khairallah}
\affiliation{Department of Physics, University of Illinois at
Urbana-Champaign, Urbana, IL 61801, USA}

\author{M.~B.~Sevryuk}
\affiliation{Institute of Energy Problems of Chemical Physics RAS,
Moscow 119334, Russia}

\author{D.~M.~Ceperley}
\affiliation{NCSA and Department of Physics, University of
Illinois at Urbana-Champaign, Urbana, IL 61801, USA}

\author{J.~P.~Toennies}
\affiliation{Max-Planck-Institut f\"ur Dynamik und
Selbstorganisation, \mbox{D-37073} G\"ottingen, Germany}

\begin{abstract}
The interplay between magic number stabilities and superfluidity
of small para-hydrogen clusters with sizes  $N = 5$ to $40$ and
temperatures $0.5~K$${} \leq T \leq 4.5~$K is explored with
classical and quantum Path Integral Monte Carlo calculations.
Clusters with $N < 26$ and T $\leq 1.5~K$  have large superfluid
fractions even  at the stable magic numbers $13$, $19$, and $23$.
In larger clusters, superfluidity is quenched especially at the
magic numbers $23$, $26$, $29$, $32$, and $37$ while below $1$~K,
superfluidity is recovered for the pairs $(27,28)$, $(30,31)$, and
$(35,36)$. For all clusters superfluidity is localized at the
surface and correlates with long exchange cycles involving loosely
bound surface molecules.
\end{abstract}

\maketitle
Hydrogen is the simplest and most ubiquitous of all molecules in
the universe. On earth, it plays an important role in many
chemical reactions and is presently being developed as an energy
transport medium \cite{http}. The $j=0$ rotational state of the
para-$H_2$ nuclear spin configuration is, like $^4He$, a
spin-less boson and below about $6~K$ has been predicted to be the
only naturally occurring superfluid besides the helium isotopes
\cite{Gin72}. The observation of superfluidity in the bulk so far
has been thwarted by its solidification at $13.96~K$. In 1991,
Sindzingre \emph{et al.} showed theoretically that small pure
clusters with $13$ and $18$ molecules were superfluid below about
$2~K$, while larger clusters with $33$ atoms had a much smaller
superfluid fraction \cite{Sin91}. Later, Grebenev \emph{et al.}
observed a superfluid response  in small clusters consisting of
$15 - 17$ para $H_2$ molecules surrounding an OCS chromophore all
within a large helium droplet \cite{Greb00}. More recently magic
cluster sizes of pure $pH_2$ clusters with $ N = 13, 33$ and $55$
were observed with Raman spectroscopy in a cryogenic free jet
expansion \cite{Tej04}. The earlier calculation of superfluidity
in pure $H_2$ clusters, as well as the interpretation of the
$OCS(pH_2)_n$ experiments have since been confirmed. Recently,
several groups have reported evidence for magic number stabilities
at $N=13$ \cite{Bar05,11,12} and at $13$ and $26$ with a reduced
superfluidity \cite{Mez0614}.

An intriguing aspect of these studies is the apparent
contradiction between the large superfluid fractions and the
structured radial distribution functions and the magic number
stabilities, which indicate a solid-like rigidity. It is only at
high temperatures, when the clusters are molten that the radial
distributions show the same constant interior density and
smooth fall-off at the surface \cite{Sin91,Schar92} found for superfluid helium
clusters which are known to be liquid \cite{Sin89}. This inconsistency has
lead to the speculation that pH2 clusters may be considered as
microscopic supersolids \cite{Sin91}. Thus the present study was undertaken
to clarify how a cluster which appears to be solid can also  be
superfluid.

To resolve this apparent incompatibility, both classical and
quantum Path Integral Monte Carlo (PIMC) calculations are reported
for all sizes between $N = 5$ and $40$ for $5$ temperatures $0.5$,
$1.0$, $1.5$, $3.0~K$ and the experimental accessible $4.5~K$
\cite{Knu96}. The calculations indicate that clusters with $N = 13$, $19$,
$23$, $26$, $29$, $32$, $34$ and $37$ have highly symmetric
structures and show a propensity for stability in agreement with
magic number stabilities reported earlier for solid ionized rare
gas clusters \cite{Far85,Mie89}. In the following, these special sizes will
be referred to as magic clusters. All the H2 clusters with $N <
26$ have significant superfluid fractions at $T\leq 1.5~K$ which is only
slightly suppressed in the magic clusters.  At magic  $N = 26$,
superfluidity is greatly quenched and for $29$, $32$, $34$ and
$37$ reduced even down to $T=0.5~K$ .  Clusters with
size pairs $(24, 25)$, $(27, 28)$, $(30, 31)$, the singleton $33$,
as well as $(35, 36)$ show  significantly greater superfluid
fractions than their more stable magic neighbours. The radial
distributions of the superfluid fraction and the distribution of
permutation cycle lengths reveal that superfluidity in all the
clusters is localized at the surface and for the larger superfluid
sizes, it correlates with the presence of loosely bound surface
molecules. These new results now clarify the apparent
contradiction between the structured radial distributions and the
large superfluidity found in the previous calculations \cite{Sin91,Mez0614}.

The calculations were carried using the PIMC method which is based
on the quantum-classical isomorphism where each particle is
replaced by a polymer made up of M ``beads''  as explained in
detail in reference \cite{Cep95}. In the present quantum calculations a
time step of $\tau = 1/80$ was sufficient to obtain converged
results within the pair-product approximation. Bose statistics is
introduced  by cross linking the polymers to form
chains of permuting cycles (polymers).  The classical
calculations involve no quantum effects such as permutations and
are described in Ref.\cite{Buch94}. Two intermolecular potentials were
used: 1) a Lennard-Jones (LJ) potential with parameters $\sigma =
2.96 \AA$  and $\epsilon = 34.16~K$ and 2) the more accurate
Silvera-Goldman (SG) potential. The initial configurations
were chosen either from the Cambridge Cluster Database \cite{con1Wal97}
or by carving out a spherical region centered around a molecule in
an hcp H2 crystal. The important effects reported here were
independent of the potential and the initial configuration.

\begin{center}
\begin{figure}[h]
\includegraphics[width=0.45\textwidth]{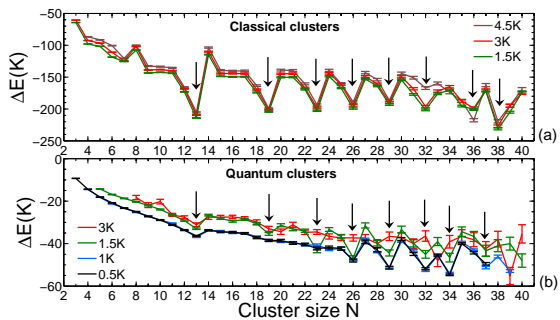}
\caption{Energy differences $\Delta E$ as a function of cluster
sizes calculated classically (a) and with a quantum mechanical
PIMC program (b). The sharp minima marked by downward arrows are
attributed to especially stable magic icosahedral-derived
structures \cite{Far85,Mie89}.} \label{Fig1}
\end{figure}
\end{center}

Magic classical and quantum clusters were identified by examining
the  energy differences $\Delta E  = E(N, T) - E(N-1,T)$, where
$E(N,T)$ is the total internal energy for a cluster of $N$
molecules at temperature $T$. $\Delta E$ approaches
the chemical potential at  $T=0$, hence sizes with $\Delta E$ values
lower than their neighbours are more localized and more tightly
bound. Both, the classical and the quantum results show about the
same enhanced relative stabilities at the magic cluster sizes $N =
13$, $19$, $23$, $26$, $29$,and $32$ indicating that the classical
stabilities persist in the quantum clusters. Although $36$ and
$38$ appear to be ``magic" in the classical simulation, the
expected magic $34$ and $37$, are found only in the
quantum calculations\cite{Far85,Mie89}. 
All our quantum calculated magic numbers
agree with well-known high symmetry icosahedral-derived structures
\cite{Far85} and have been repeatedly observed in ionized rare gas
clusters \cite{Far85,Mie89}.  So far, however, only magic $N = 13$, $33$ and
$55$, have been observed experimentally \cite{Tej04}. Whereas $N = 13$ and
$55$ correspond to the closing of the first and second icosahedral
shells, $N = 33$ is not a magic number among the possible modified
icosahedral structures but has a dodecahedral form. Since the
experimental resolution \cite{Tej04} was not sufficient to distinguish from the
nearby magic $32$ and $34$, it is possible that the
earlier assignment may be incorrect. Especially if one considers
that during cluster growth icosahedral modified structures will be
preferred as a result of build-up around the smallest magic $N=13$
which will be a nucleus for further growth.

%\begin{center}
\begin{figure}[h]
\includegraphics[width=0.45\textwidth]{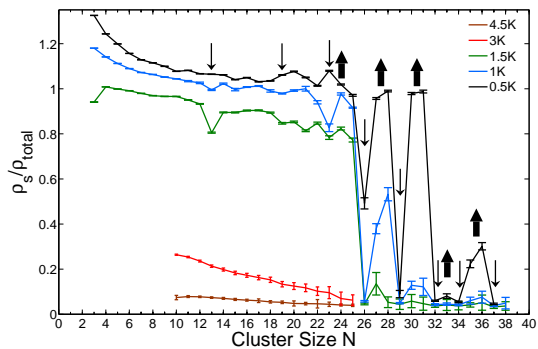}
\caption{The total superfluid density fraction is plotted as a
function of cluster size for five different temperatures. The
downward arrows indicate the magic clusters with high symmetry.
The upward pointing arrows indicate islands of sizes exhibiting
enhanced superfluidity. Superfluidity beyond $N=26$ above $3~K$ is
below $3\%$.} \label{fig2}
\end{figure}
%\end{center}

The superfluid fraction $\rho_s/\rho_{total}$ \cite{Cep95}
is shown in Fig.\ref{fig2} as a function of cluster size and
temperature. The nearly $100\%$ fraction in the small
clusters decreases sharply above $1.5~ K$ \cite{24}. At each of the
magic sizes $\rho_s/\rho_{tot}$ almost
always shows a downward dip reminiscent of those in Fig.
1. These dips are smallest in the small clusters but are more
pronounced for $N \ge 26$. Thus, the classical rigidity at the
magic sizes suppresses the quantum delocalization needed for
superfluidity, an effect which becomes stronger with increasing
size. Surprisingly, beyond  magic  $N = 23$, the superfluid
fraction at $1~K$ jumps back to about unity for the next larger
sizes $(24,25)$. Then beyond magic $N = 26$ a similar
rebound occurs for the two lowest temperatures at $(27,28)$. 
The minimum at $N = 23$ and two maxima at $N = 25,27$
were also found in recent calculations by Mezzacapo and Bonisegni
\cite{Mez0614}. In the present calculations similar maxima are found at
$(30,31)$,and $(35,36)$. The singleton $N=33$
has a very weak rebound and seems to follow the behaviour of the
neighbouring magic sizes as its relatively low binding energy in
Fig.1(b) also suggests. Thus in clusters with solid-like stable
structures superfluidity is suppressed, with the largest
suppressions found for $N\ge 26$, while at sizes
not corresponding to known magic sizes, it is restored.
It is interesting to note that at $1.5~ K$ the interactions are so
strong that superfluidity is almost suppressed for clusters $N
\ge 26$, yet quantum delocalization in the smaller clusters is
still sufficient for their superfluidity.

%\begin{center}
\begin{figure}[h]
\includegraphics[width=0.5\textwidth]{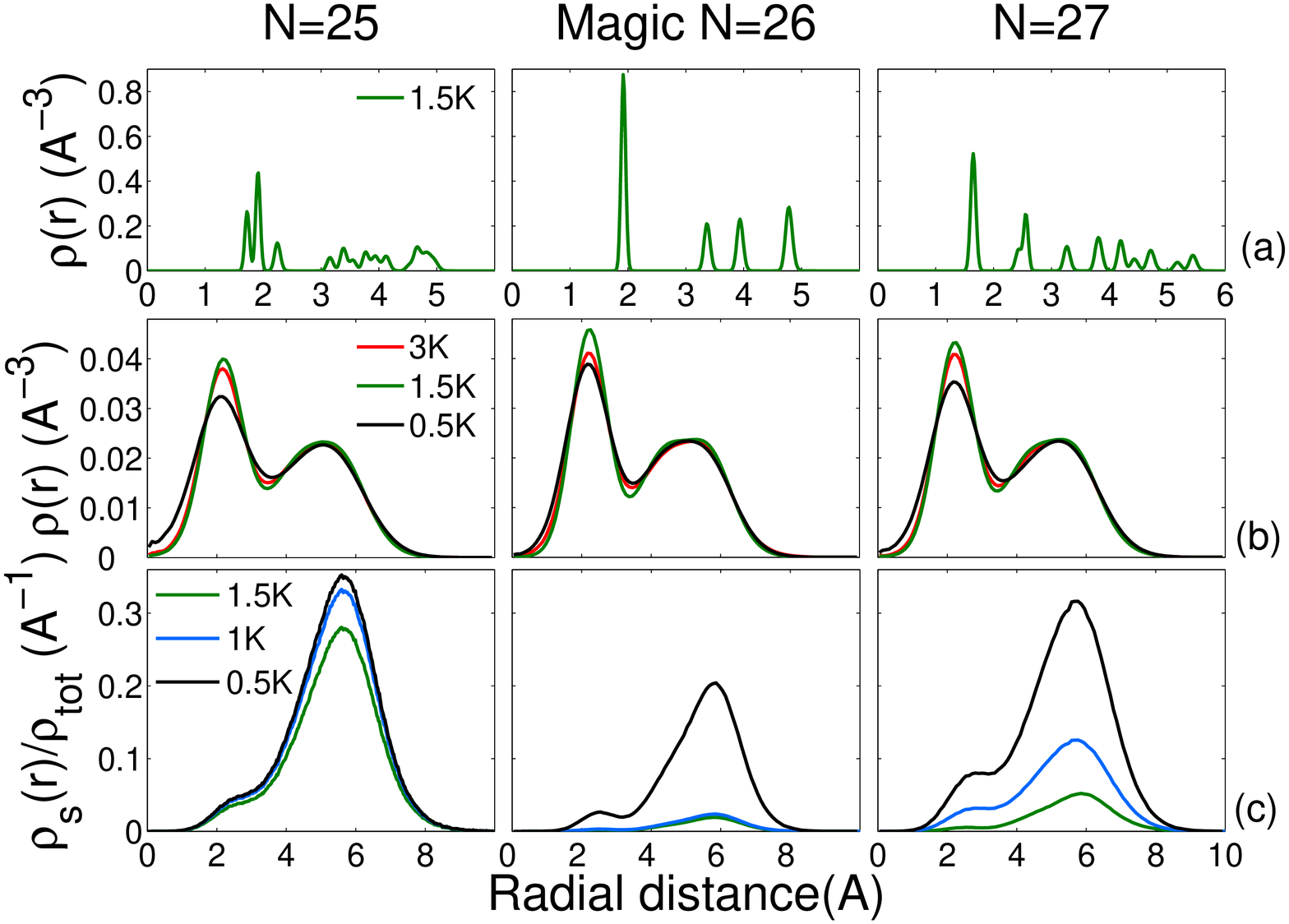}
\caption{Comparison of radial density distributions for the magic
cluster $N = 26$ and the neighbours $(25, 27)$
for three characteristic
temperatures. The distributions in (a) are classical and those in
(b) are from quantum calculations. The radial distribution of the
superfluid fraction is plotted in (c). We show one temperature in (a) 
since the classical 
distributions are less sensitive to temperature. } \label{fig3}
\end{figure}
%\end{center}

To understand these unexpected out-of-phase oscilllations between
magic sizes and superfluid sizes the three types of radial
distributions shown in Fig.3 were calculated for the magic $N =
26$ cluster and the two adjacent less stable but more superfluid
clusters. Fig.3 (a) shows the classical
distributions at $1.5~ K$. As expected classicaly(Fig.3(a)),the molecules 
in $N=26$ follow an orderly partitioning into
four localised and distinct groups, whereas for $N = 25, 27$ they
are randomly distributed. The quantum distributions
(Fig. 3b) with only two broad peaks show evidence of quantum
exchanges between the shells. %agree very well with those recently
%published [11-14]. 
%The much greater smearing with only two broad
%peaks in comparison to the classical distributions is a dramatic
%illustration of quantum melting discussed recently in connection
%with a temperature reentrant behavior[10-13]. 
The magic $N = 26$
cluster has a noticeably larger inner peak compared to its
neighbors and also shows a slight dimple in the second outermost
peak reminiscent of the classical distribution. The radial
dependence of the  superfluid fraction $\rho_s(r)/\rho$ \cite{25} 
in Fig.3(c) is computed by
binning the radial location of the beads that are involved in
permutation cycles \cite{25}. These distributions 
exhibit large
differences with temperature and with sizes as expected from Fig.
2. Surprisingly, however, superfluidity is not greatest
near the center as found for $^4$He clusters \cite{Sin89}, but is 
localized at the surface
beyond the outer maximum in the radial density distribution. The
superfluidity is small in the inner shell specially for magic $N=26$. 
We note that the
apparent randomness in the classical radial curves
(Fig.3(a)), which is suggestive of less rigid and symmetric
structures for the non-magic $N=25$ and $27$, correlates with the
increased superfluid fractions (Fig. 3(c) and Fig. 2).

%\begin{center}
\begin{figure}[h]
\includegraphics[width=0.5\textwidth]{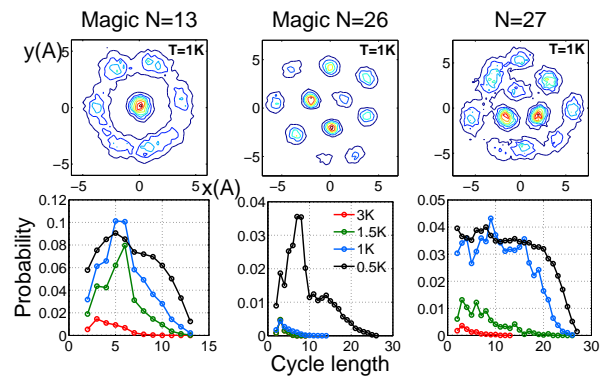}
\caption{ Top panels: Density contour plots at $z\approx 0$
plane for magic $N=13$, $26$ and non-magic $N=27$. Bottom
panels:The probability distributions for permutations of a given
length.  The contours for
magic $N = 13$, show delocalized molecules at the outer layer
that take part in cycles with lengths $5$ and $6$ which
correspond to rings around the central molecule. In magic $N=26$
at $1.0~ K$, all the molecules are highly localized. The surface
molecules in $N = 27$ cluster at $1~ K$ are highly
delocalized, whereas the core molecules are nearly rigid. This
correlates with the long exchange paths around the localized core
shown below the contour plot.} \label{fig4}
\end{figure}
%\end{center}

Additional insight comes from Fig.4 which
provides a cross section view of
$N=13$, $26$ and $27$ and the probability distributions
of the permutation cycles as a function of the permutation lengths. 
The contours for $N=13$ show
a considerable delocalization of the outer layer. The
$N=13$ permutation probabilities are dominated by cycles with
lengths of  $5$ and $6$ molecules, corresponding to rings around the
center, with a relatively high probability of $10\%$ in
accordance with the large superfluid fraction (Fig. 2). Even
though the central molecule appears to be localized in the contour
plot at $0.5~K$, it also  participates in  ring exchanges as
indicated by the small $\approx 0.1\%$ probability for cycle
exchange lengths of $13$. The contour plot for magic $N=26$
indicates that all its molecules are highly localized and that it
is solid-like justifying the greatly suppressed superfluidity
(Fig. 2).  Contours at different values of $z$ 
for larger clusters  at $T \geq 1.5 ~K$ 
(not shown) show similar localization. The corresponding
permutation probabilities are peaked at very low cycle numbers and
are mostly less than $0.05\%$, which explains their small residual
superfluidity in Fig. 2. But as shown by the permutation 
probabilities even the magic
$N=26$ melts and become superfluid at $0.5~ K$ with a predominance of
cycles from $7$ to $15$. The top panel in Fig. 4 for  non-magic
$N=27$ has a liquid outer layer similar to magic $13$ in agreement
with Fig. 3(c) but with a smaller peak permutation probability of
about $3.5\%$ as expected from its small overall superfluid
fraction (Fig. 2). At $0.5~K$ the permutation probability curve
smoothens and extends out to include cycles equal to
$27$ suggesting that also the core molecules are
participating in the permutations. Thus both the contour
plots and probability distributions of the $N=27$ cluster are
consistent with the onset of superfluidity in the surface region.

Recently, Mezzacapo and Boninsegni also
observed an enhanced superfluidity for $N = 25$, but their
conclusion that ``the addition of a molecule to the $N+1= 26$ has
the effect of frustrating the solid order of the inner shell,
increasing molecule delocalization and leading to quantum
exchanges'' \cite{Mez0614} is at variance with these new results, which
clearly show that superfluidity is at the surface.

The nature of the disorder favouring a large superfluid fraction
at the surface emerges from the ``inherent structure'' (IS)
analysis of Stillinger and Weber \cite{Web84}: ``inherent structures
which underlie the liquid state are those stable particle packings
(potential minima) which can be reached by  a steepest descent
quench on the potential energy hypersurface''. This quenching
procedure eliminates all kinetic effects due to thermal
excitations or zero-point motion. To generalize this concept to
quantum mechanical systems, in the steepest descent minimization,
the gradient was calculated from the path
integral action rather than from the potential.
The permutations were turned off and the IS analysis
was applied to ``Boltzmannons". The quenched structures are, in
general, independent of the temperature \cite{Web84} and are identical
to Wales classical minimum potential energy configurations
\cite{con1Wal97}. This geometric spatial correspondence explains the origin
of the persistence of the classical structures in the energy and
superfluid densities discussed above.

Among the small cluster sizes, a quenched configuration similar to
the Wales body centered icosahedron was quite often found for $N =
13$ indicating that it is particularly stable. Occasionally,
quenching would generate variant structures close to the classical
ones with some delocalized molecules on the surface suggestive of
melting. Clusters differing by one or two molecules from the magic
sizes are more often seen to have defective surfaces. Since these
clusters differ essentially only in the bonding of the outer
molecules, their smaller binding energies (Fig.2) indicate that
these outer molecules are less tightly bound than in the case of
the magic clusters. For example, $N = 18$ would statistically
appear more often with structures that deviate slightly from the
classical clusters, while for magic $N = 19$,  almost every IS
cluster is the same as the classical cluster. The IS analysis for
the other larger superfluid sizes indicates that their surface
molecules are also less tightly bound and less ordered.

In summary, our analysis reveals that pure pH2 clusters
with $N < 26$ at temperatures $T \leq 1.5 ~K$ are liquid-like
and have a large superfluid response. It is only somewhat reduced
in magic clusters $N = 13,19$ and $23$, which are classical
magic sizes with highly symmetric icosahedral structures.
According to Fig 1 (b), the difference in internal energies, which
is the energy needed to add one molecule, is less
than $38 ~K$ for these highly superfluid clusters. The larger
magic clusters $N=26, 29, 32, 34$, and $37$, in which
superfluidity is strongly quenched at temperatures $T>0.5~ K$,
all have considerably larger internal energy differences of more
than about $40~ K$. Superfluidity is restored in the cluster size
pairs $N =(24,25)$, $(27,28)$, $(30,31)$, the singleton
$33$, and $(35,36)$, with smaller internal energy differences
compared to the magic clusters which lie in between. In these
superfluid sizes, quantum delocalization  of the loosely bound
admolecules enables them to explore many different surface
structures, thereby favouring large permutation cycles and an
increased superfluidity. Our calculations reveal that
with increasing cluster size the strong many-body intermolecular
interactions lead to a rigid solid-like inner core thereby pushing
the delocalization induced superfluidity towards the surface,
where it is favored by the reduced coordination and weak inward
interaction with the small central core. The overall decay of
superfluidity with cluster size and its increased localization on
the surface agrees with the macroscopic limit of 2D surface
superfluidity and zero response in the bulk \cite{27}.

We thank Dr.Oleg A. Kornilov and Prof. L. Yu
Rusin for helpful discussions and Prof. Victoria Buch for her
classical code. This work was supported by
NSF(DMR-04-04853), NASA (NAG-8-1760) and 
the Deutsce-Forschungsgemeinschaft. Computer time
was provided by NCSA, the F. Seitz Materials Research Lab.
(US DOE DEFG02-91ER45439 and NSF DMR-03 25939 ITR) at the U. of
Illinois Urbana-Champaign.


\begin{thebibliography}{99}

\bibitem{http}http://www.eere.energy.gov/hydrogenandfuelcells/
\bibitem{Gin72}V. L. Ginzburg and A. A. Sobyanin, JETPLetters {\bf 15}(6) 343 (1972).
\bibitem{Sin91} P. Sindzingre, D. M. Ceperley, and M. L. Klein,
  Phys. Rev. Lett., {\bf 67}, 1871-1874 (1991)

\bibitem{Greb00}S. Grebenev, B. Sartakov, J. P. Toennies, 
A. F. Vilesov, Science, {\bf 289}(5484), 1532 - 1535 (2000).
%\bibitem{Sin91}P. Sindzinge, D. M. Ceperley, and M. L. Klein,
%  Phys. Rev. Lett. {\bf 67} (14), 1871-1874 (1991)

\bibitem{Tej04} G. Tejeda J. M. Fern\'andes, S. Montero, D. Blume,
J. P. Toennies, Phys. Rev. Lett., {\bf 92}(22), 223401 (2004).

\bibitem{Schar92} D. Scharf, G. J. Martyna, M. L. Klein,
  Chem. Phys. Lett., {\bf 197}(3), 231 - 235 (1992). %. Scharf, M. L. Klein and G. J. Martyna, J. Chem.  Phys. 97, 3592 (1992)

\bibitem{Kwo02}Y. Kwon, K. B. Whaley, Phys. Rev. Lett., {\bf
    89}, 273401 - 1 (2002).

%\bibitem{9} M. A. McMahon and K. B. Whaley, Chem. Phys. 182, 119
%(1994).

\bibitem{Bar05}S. Baroni, S. Moroni, Chem. Phys. Chem., {\bf 6}(9),
1884-1888 (2005).

\bibitem{11}J. E. Cuervo, P. N. Roy, J. Chem. Phys. 125, 124314
(2006).


\bibitem{12}R. Guardiola, J. Navarro. Phys. Rev. A74, 025201
(2006).

\bibitem{Mez0614}F. Mezzacapo, M. Boninsegni, Phys. Rev. Lett., {\bf
    97}(4), 045301 (2006). F. Mezzacapo, M. Boninsegni, cond-mat / 0611775.

%\bibitem{14}F. Mezzacapo and M. Boninsegni, cond-mat / 0611775

\bibitem{Sin89} P. Sindzingre, M. L. Klein, and D. M. Ceperley,
  Phys. Rev. Lett., {\bf 63}, 1601 (1989)

\bibitem{Knu96}E. L. Knuth, F. Schünemann, and J. P. Toennies,
  J. Chem. Phys. {\bf 102}, 6258 (1995).

%\bibitem{Buch93}V. Buch and J. P. Devlin, J. Chem. Phys., {\bf
%    98}(5), 4195 - 4206 (1993).
\bibitem{Far85} J. Farges, M. F. de Ferandy, B. Raoult and G. Torchet,
  Surface Science, {\bf 156}, 370 (1985).


\bibitem{Mie89}W. Miehle, O. Kandler, T. Leisner and O. Echt,
  J. Chem. Phys., {\bf 91}, 5940 (1989).

\bibitem{Cep95}D. M. Ceperley,Rev.Modern Phys.,{\bf 67}(2),279-355 (1995).

\bibitem{Buch94}V. Buch, J.Chem.Phys.,{\bf100}(10),7610-7629 (1994).

%\bibitem{Sil78}I. F. Silvera and V. V. Goldman, J. Chem. Phys., {\bf
%    69}(9), 4209 - 4213 (1978).

\bibitem{con1Wal97} http://www-doye.ch.cam.ac.uk/jon/structures
/LJ/tables.150.html. D. J. Wales, J. P. Doye, J.Phys.Chem. A,{\bf 101}(28),5111-5116 (1997).

%\bibitem{Wal97}D.J.Wales and J.P.Doye, J.Phys.Chem. A,{\bf 101}(28),5111-5116 (1997).

\bibitem{24}Values greater than 100$\%$ in Fig.2 result from 
ambiguity in the definition of the moments of inertia in small clusters.

\bibitem{25}E. W. Draeger, D. M. Ceperley, Phys.Rev.Lett.90, 065301 2003.

\bibitem{Web84}T. A. Weber, F. H. Stillinger, J. Chem. Phys., {\bf 81}(11), 5089-5094 (1984). 

\bibitem{27}M. Wagner, D. M. Ceperley, J. Low Temp. Phys. 94, 147 (1994)


\end{thebibliography}
\end{document}